\documentclass[aps,prd,superscriptaddress,onecolumn,12pt,tightenlines]{revtex4}

\usepackage{graphicx}
\usepackage{ifthen}

\def\gae{\lower 2pt \hbox{
 $\, \buildrel {\scriptstyle >}\over {\scriptstyle \sim}\,$}}
\def\lae{\lower 2pt \hbox{
 $\, \buildrel {\scriptstyle <}\over {\scriptstyle \sim}\,$}}
\newcommand{\ifonecolumn}[2]{
  \ifthenelse{\lengthtest{\columnwidth=\textwidth}}{#1}{#2}
}
\newcommand{\greaterofdim}[2]{\ifdim#1>#2#1\else#2\fi}
\newcommand{\lesserofdim}[2]{\ifdim#1<#2#1\else#2\fi}

\begin{document}


\preprint{MCTP-04-14}


\title{Age of the Universe in Cardassian Models}


\author{Chris Savage}
\affiliation{
 Michigan Center for Theoretical Physics,
 Physics Department,
 University of Michigan,
 Ann Arbor, MI 48109}

\author{Noriyuki Sugiyama}
\affiliation{
 Michigan Center for Theoretical Physics,
 Physics Department,
 University of Michigan,
 Ann Arbor, MI 48109}
\affiliation{
 Department of Physics,
 University of Wisconsin -- Milwaukee,
 Milwaukee, WI 53211}

\author{Katherine Freese}
\affiliation{
 Michigan Center for Theoretical Physics,
 Physics Department,
 University of Michigan,
 Ann Arbor, MI 48109}

\date{\today}


\begin{abstract}
  The age of the universe is obtained in a subset of Cardassian models
  by using WMAP data.  Cardassian expansion is a modification to the
  Friedmann equation that allows the universe to be flat, matter
  dominated, and accelerating, without a vacuum component. Since this
  model changes the evolution of the universe, we should not \textit{a
    priori} expect the Cardassian age to be the same as the WMAP
  Friedmann derived result of $13.7 \pm 0.2$ Gyrs.  However, in the
  subset of Cardassian models we consider, we discover that the age of
  the universe varies from 13.4-13.8 Gyr over the range of parameter
  space we explore, a result close to that of the standard $\Lambda$
  Cold Dark Matter model.  
  The Hubble constant $h$, which may also vary in these
  models, likewise varies little from the Friedmann result.
\end{abstract}

\maketitle


\section{\label{sec:Intro} Introduction}

Over the past five years, Cosmic Microwave Background (CMB) data
have shown the universe to be flat
\cite{omega1, wmap} and supernova data have indicated that it is
accelerating \cite{SNCP:HighZ,SNST:Accel}.  The matter
content of the universe falls well short of the necessary energy
density to provide a flat curvature in the standard cosmological
model, the Friedmann model.  The most popular interpretation of
this mismatch is that
the ``missing'' density is assumed to be
present in the form of a vacuum (or dark) energy that provides a
pressure leading to the acceleration of the universe.  

Alternatively, general relativity may need to be modified.
Modification of
the standard cosmological model, as in the Cardassian model
\cite{Freese:2002sq}, allows both the flatness and acceleration to
be accounted for solely due to the matter content of the universe.
This modification changes the evolution of the universe.  Hence the
CMB anisotropy spectrum will imply a
different age of the universe \cite{wmap}.

The Cardassian model \cite{Freese:2002sq,Freese:2002gv,Gondolo:2002fh,
Wang:2003cs,Sen:2003cy} 
modifies the Friedman equation by adjusting the right hand side to be a
more general function of the energy density.  This function returns to
the usual Friedmann equation for the early history of the universe, so
that ordinary nucleosynthesis takes place. However, in the
recent past, beginning at a redshift $O(1)$, this function drives the
universe to accelerate.  Such modifications to the Friedmann equation
may arise if our universe exists on a brane embedded in a higher
dimensional universe \cite{Chung:XDimensions}. Other proposed
modifications to the Friedmann equations include \cite{other}.

In section~\ref{sec:Friedmann}, we
will review the Friedmann model and describe how to determine the age
of the universe.  In section~\ref{sec:WMAP}, Wilkinson Microwave
Anisotropy Probe (WMAP) results are reviewed and the choice of
appropriate parameters is discussed.  In section~\ref{sec:Cardassian},
we will expand these techniques to the Cardassian model.  
We note that, since the
customary parameters $h$, $\Omega_b$, and $\Omega_m$ are not orthogonal
and indeed have values that depend upon the cosmological model itself,
more suitable parameters must be extracted from the data in order
to obtain the universe age in any model.
And finally,
results for the Cardassian model are discussed in section
\ref{sec:Results}. In this paper, we obtain the age of the universe
only for the Modified Polytropic (MP) Cardassian models defined below.

\section{\label{sec:Friedmann} Age in the Friedmann Model}

In standard cosmology, the
evolution of the universe is governed by the Friedmann equation:
\begin{equation} \label{eqn:Friedmann}
 H^2 = \frac{8 \pi G \rho}{3}
\end{equation}
where $H$ is the Hubble parameter and we have dropped the curvature
term as experimental results are consistent with a flat universe
\cite{omega1, wmap}.  At the curent epoch the critical density is
\begin{equation} \label{eqn:CritDensity}
 \rho_c = \frac{3 H_0^2}{8 \pi G}
        = 1.054 \times 10^{-5} h^2 \ \textrm{GeV}/\textrm{cm}^3
\end{equation}
where subscript $0$ refers to the present day and
$H_0 = 100 h~\textrm{km/s/Mpc}$.
Here, $\Omega \equiv \frac{\rho}{\rho_c}$ is the ratio of energy
density to the critical density, with $\Omega_i$ representing
the ratio due to the component $\rho_i$ of the density.
In the standard  picture, an additional component beyond matter and
radiation is assumed to reach the critical density. This component
is taken to
be a vacuum energy: a cosmological constant $\Lambda$ 
or a time dependent vacuum
energy, or scalar field known as ``quintessence'', that evolves
dynamically with time \cite{yaya}.

The Hubble parameter can be related to its present day value by:
\begin{eqnarray}
 \label{eqn:HubbleEvol}
  H &=& H_0 E_F(z) \\
 \label{eqn:HubbleE}
  E_F(z)^2 &=& \Omega_{0r} (1+z)^4 + \Omega_{0m} (1+z)^3
               \ifonecolumn{}{\nonumber\\ &{}&{} \qquad}
               + \Omega_{0X} (1+z)^{3(1+w_X)}
\end{eqnarray}
where $\Omega_{0r}$, $\Omega_{0m}$, and $\Omega_{0X}$ are the current
contributions from radiation, matter, and vacuum respectively, and
$w_X=p_X/\rho_X$ is the equation of state.

To determine the age of the universe, we integrate Eqn.\ 
(\ref{eqn:Friedmann}), 
\begin{equation} 
\label{eqn:Age}
 t_0 = {H_0}^{-1} \int_{0}^{\infty} \frac{dz}{(1+z)}
       \frac{1}{E_F(z)} \, .
\end{equation}

\section{\label{sec:WMAP} WMAP Parameters}

The WMAP Collaboration has determined values for several parameters by
mapping the cosmic microwave background \cite{wmap};
relevant ones are listed in Table~\ref{table:Parameters}.  They have
determined the age of the universe in a $\Lambda$-CDM (cold dark matter)
 model which assumes that
the dark energy is a cosmological constant,
\begin{equation}
 t_0 = 13.7 \pm 0.2 \ \textrm{Gyr}
       \quad\ (\Lambda{\rm -CDM}\, \ \textrm{model})  ,
\end{equation}
where the uncertainty is determined by statistical analysis of their
Monte Carlo Markov Chain results \cite{WMAP:Methodology}.

\begin{table}    
  \centering
  \begin{tabular*}%
      {\lesserofdim{0.60\textwidth}{1.00\columnwidth}}%
      { l @{\extracolsep{\fill}} c @{\extracolsep{0 pt}} @{ = } c }
    \hline \hline
    \multicolumn{3}{c}{WMAP Parameters} \\[0.5 ex]
    \hline
    \noalign{\vspace{0.5 ex}}
    Hubble constant          & $h$           & $0.71_{-0.03}^{+0.04}$    \\
    baryon density           & $\Omega_{0b}$ & $0.044 \pm 0.004$         \\
                             & $\omega_b$    & $0.0224 \pm 0.0009$       \\
    matter density           & $\Omega_{0m}$ & $0.27 \pm 0.04$           \\
                             & $\omega_m$    & $0.135_{-0.009}^{+0.008}$ \\
    acoustic scale           & $\ell_A$      & $301 \pm 1$               \\
    redshift of decoupling   & $z_{dec}$     & $1089 \pm 1$       \\[0.5 ex]
    \hline \hline
  \end{tabular*}
  \caption[WMAP Parameters]{
    Cosmological parameters determined by the WMAP Collaboration from
    fits to multiple experiments \cite{wmap}.
  }
  \label{table:Parameters}
\end{table}

Using the WMAP results for $h$ and $\Omega_{0m}$ in Eqn.\ 
(\ref{eqn:Age}) yields the correct value, but overestimates
the uncertainty.  This is due to the fact that the parameters $h$
and $\Omega_{0m}$ are not orthogonal; that is, their uncertainties
are correlated \cite{Percival:Constraints, WMAP:Peaks}.
Instead, the set of (nearly) orthogonal parameters are $\omega_m
\equiv \Omega_{0m} h^2$, $\omega_b \equiv \Omega_{0b} h^2$ (where $b$
is for baryon), and the acoustic scale $\ell_A$.

Oscillations in the photon-baryon fluid during the early universe led
to peaks in the CMB power spectrum.
The first peak in this spectrum is related to the angle
$\theta_A$ subtended by the conformal distance $s$ a sound wave
travelled from the big bang until decoupling at the surface of last
scattering, which has a conformal distance denoted by $D$.  In a flat
universe,
\begin{equation} \label{eqn:AcousticScale}
 \theta_A = \frac{\pi}{\ell_A} = \frac{s}{D}
\end{equation}
The distances $s$ and $D$ are given by:
\begin{eqnarray}
 \label{eqn:SoundHorizon}
  s &=& \frac{1}{H_0} \int_{z_{dec}}^{\infty} dz \, \frac{c_s}{E_F(z)} \\
 \label{eqn:DistanceLS}
  D &=& \frac{1}{H_0} \int_{0}^{z_{dec}} dz \, \frac{1}{E_F(z)}
\end{eqnarray}
where $c_s$ is the speed of sound in the photon-baryon fluid:
\begin{equation} \label{eqn:SoundSpeed}
 c_s = \frac{1}{\sqrt{3 \left( 1
       + \frac{3}{4} \frac{\rho_b}{\rho_{\gamma}}\right) }}
\end{equation}

An in-depth discussion of a photon-baryon fluid and the CMB may be found
in Refs.~\cite{Hu:1996qs,Hu:1995en,Hu:1994jd,Hu:2000ti,Tegmark:1995kq,
Knox:2001fz}. 

The three parameters $\omega_m$, $\omega_b$, and $\ell_A$ have nearly
orthogonal effects on the CMB power spectrum; $\ell_A$ is related to
the position of the first peak, while the other two are related to the
peak heights.

The redshift of photon-baryon decoupling $z_{dec}$ is determined
mainly from the overall temperature of the CMB,
with a WMAP value of $1089 \pm 1$.  The redshift of matter-radiation
equality $z_{eq}$ is given by:
\begin{equation} \label{eqn:ZEq}
 1 + z_{eq} = \frac{5464}{1 + \rho_{\nu} / \rho_{\gamma}}
              \frac{\omega_m}{0.135}
              \left( \frac{T}{2.725 \, \textrm{K}} \right)^4
\end{equation}
with a neutrino to photon density ratio of $\rho_{\nu} / \rho_{\gamma}
= 0.6851$ and an overall CMB temperature of $T = 2.725 \, \textrm{K}$
\cite{WMAP:Peaks}.

While several of the above equations may be greatly simplified in the
standard Friedmann model 
(with closed forms available for some of the integrals)
\cite{WMAP:Peaks}, we will not make these simplifications here.
Instead, it will be useful to use numerical techniques that can be
applied below to the Cardassian
Model, where the equivalent integrals
do not have closed forms.  From Eqn.~(\ref{eqn:AcousticScale}),
we can take $h$ to be an implicit function of $\omega_m$, $\omega_b$,
and $\ell_A$:
\begin{equation} \label{eqn:hFct}
 h \equiv h ( \omega_m, \omega_b, \ell_A )
\end{equation}
That is, given observed values for $\omega_m$, $\omega_b$, and
$\ell_A$ (from WMAP), we can use numerical routines to find the value
of $h$ necessary to satisfy $\ell_A = \frac{\pi D}{s}$, where $s$ and
$D$ (dependent upon $\omega_m$, $\omega_b$, and $h$ only) are given
by Eqns. (\ref{eqn:SoundHorizon}) \& (\ref{eqn:DistanceLS}),
respectively.  The value of $h$ thus obtained may then be inserted
into the time integral, Eqn.~(\ref{eqn:Age}) (using
$\Omega_{0m} = \omega_m / h^2$), to obtain an age that is dependent
on the three orthogonal parameters.

This procedure correctly reproduces both the Hubble constant $h$ and
age of the universe $t_0$ (including uncertainties) obtained through
the WMAP MCMC, listed in Table~\ref{table:Parameters}.
Reproduction of the correct uncertainties affirms that the
aforementioned parameters are indeed nearly orthogonal and will be
the appropriate parameters to use in more general models.

\section{\label{sec:Cardassian} Age in Cardassian Models}

Cardassian expansion was proposed as a model in which matter alone
is sufficient to drive acceleration of the universe; in this 
model, there is \textit{no} vacuum energy.  Instead, the Friedmann
equation is modified to give
a general function of the energy density on the right hand side:
\begin{equation} \label{eqn:GeneralFriedmann}
 H^2 = g(\rho)
\end{equation}
This function returns to the usual Friedmann equation
for the early history of the universe, so that ordinary
nucleosynthesis and evolution results. However, in the recent
past, beginning at a redshift $O(1)$, this function drives
the universe to accelerate.

We see the critical density $\tilde{\rho}_c$ is then defined by the
relation:
\begin{equation} \label{eqn:GenCritDensity}
 g(\tilde{\rho}_c) = H_0^2
\end{equation}
where we note the critical density is not, in general, the same as in
the original Friedmann model (Eqn.~(\ref{eqn:CritDensity})).  It will
be useful to relate the new critical density and other parameters to
the old critical density.  To prevent confusion, we will use
$\tilde{\rho}_c$ to denote the Cardassian critical density, while
$\rho_{c}$ will continue to denote the Friedmann critical density.
Likewise, $\Omega_i \equiv \frac{\rho_i}{\rho_c}$ will continue to be
defined in terms of the Friedmann critical density.

In this section, we will look at a subset of the Cardassian
models \cite{Freese:2002sq, Freese:2002gv}
known as Modified Polytropic Cardassian (MP Cardassian):
\begin{equation} \label{eqn:Cardassian}
 H^2 = g(\rho)
     = \frac{8 \pi G \rho}{3}
             \left[ 1 + \left( \frac{\rho}{\rho_{card}}
                    \right)^{q (n - 1)}
         \right]^{1/q}
\end{equation}
where $\rho_{card}$, $q \geq 1$, and $n <2/3$ are free
parameters.  One of these parameters is fixed by the observed value
of $\Omega_{m0}$, and a constraint on the remaining two
is obtained by matching supernova data (the
redshift at which the second term becomes important).
The normalization is fixed to match the Friedmann model at early
times.

We
will require $n < 2/3$, so that the second term in the
brackets of Eqn.~(\ref{eqn:Cardassian}) increases over time,
eventually dominates, and then leads to acceleration of the universal
expansion.  In this paper we consider only $n\geq 0$; in fact
it could be negative as well. The case $q=1$
is equivalent to the original power law form of the Cardassian
model \cite{Freese:2002sq}, and the case $q = 1$ \& $n = 0$ is
equivalent to $\Lambda$-CDM with the second term 
representing the vacuum component.  

Taking the geometry to be flat, the critical density condition
Eqn.~(\ref{eqn:GenCritDensity}) gives:
\begin{equation} \label{eqn:CardCritDensity}
 \tilde{\rho}_c
          = \rho_{c}
            \left[
              1 + \left( \frac{\rho_0}{\rho_{card}} \right)^{q (n-1)}
            \right]^{\frac{-1}{q}}
\end{equation}
Here the critical density can be smaller than in the original
Friedmann case. We will choose the parameters so that matter alone is
sufficient to provide today's critical density, without any vacuum
contribution, $\rho_{0m} \approx \rho_0 = \tilde{\rho}_c$, so that
\begin{equation} \label{eqn:CardOmegaM}
 \Omega_{0m} \equiv \frac{\rho_{0m}}{\rho_{c}}
 = \frac{\tilde{\rho}_c}{\rho_{c}} .
\end{equation}
The numerical value of $\Omega_{0m}$ will be roughly 1/3 when matching
this model to the CMB data (see the discussion in the results
section below).  The condition in Eqn.~(\ref{eqn:CardOmegaM}) is
satisfied for parameters satisfying
\begin{equation} \label{eqn:CardParam}
 \rho_{card} = \rho_0 \left[ \Omega_{0m}^{-q} - 1
                      \right]^{\frac{1}{q (1 - n)}} ,
\end{equation}
or, equivalently,
\begin{equation} \label{eqn:CardRedshift}
 1 + z_{card} = \left[ \Omega_{0m}^{-q} - 1
                \right]^{\frac{1}{3 q (1 - n)}} .
\end{equation}
After this redshift, the second term on the RHS of Eqn.
(\ref{eqn:Cardassian}) begins to dominate and (for $n<2/3$) the
universal expansion will accelerate (even without a vacuum component).

With the previous constraints, we may rewrite Eqn.
(\ref{eqn:Cardassian}) as:
\begin{eqnarray}
 \label{eqn:CHubbleEvol}
  H &=& H_0 E_C(z,q,n) \\
 \label{eqn:CHubbleE}
  E_C(z)^2 &=& \Omega_{0r} (1+z)^4 + \Omega_{0m} (1+z)^3
               \ifonecolumn{}{\nonumber\\ &{}&{} \qquad}
             + \Omega_{0X} f_X(z)
\end{eqnarray}
where (treating $\Omega_{0r}$ as negligible):
\begin{eqnarray}
 \label{eqn:COmegaX}
  \Omega_{0X} &=& 1 - \Omega_{0m} \\
 \label{eqn:CfX}
  f_X(z) &\equiv& \frac{(1 + z)^3}{\Omega_{0m}^{-1} - 1}
                  \bigg\{
                    \left[
                      1 + (\Omega_{0m}^{-q} - 1) (1 + z)^{3 q (n-1)}
                    \right]^{1/q}
                  \ifonecolumn{}{\nonumber\\ &{}&{} \qquad\qquad\qquad}
                    - 1
                  \bigg\}
\end{eqnarray}
and we now have only two free parameters, $q$ and $n$.

Note that at high redshifts $z$, $E_C(z,q,n) \to E_F(z)$
for all allowable choices of $q$ and $n$, so that at early times,
$E_C(z,q,n)$ is equivalent to the Friedmann case (compare Eqn.
(\ref{eqn:HubbleEvol}) to Eqn.~(\ref{eqn:CHubbleEvol})).
Consequently, the evolution of the universe before and around the
period of decoupling ($z > 1000$) is the same in the Cardassian
model as in the Friedmann model so that ordinary nucleosynthesis
and oscillations in the photon-baryon fluid are unaffected.

The age of the universe may be derived in the same manner as in
the standard Friedmann model, and we find
\begin{equation} \label{eqn:CardAge}
 t_0 = {H_0}^{-1} \int_{0}^{\infty} \frac{dz}{(1+z)}
         \frac{1}{E_C(z,q,n)} \, .
\end{equation}
This integral does not in general have a closed form and must be
evaluated numerically.

Note the CMB peak amplitudes depend upon the parameters $\omega_m$ and
$\omega_b$, but are essentially independent of the late time expansion.
Thus, the WMAP fits to these two parameters are valid-- the Cardassian
model does not affect their determination from these peak amplitudes.
The values for $\Omega_{0m}$ and $\Omega_{0b}$ are not fixed, however,
as $h$ is not independently determined from the amplitudes.  Instead,
$h$ may be determined from the peak positions, expressed by $\ell_A$,
which also depends upon the expansion model.  In our analysis, we may
fix $\omega_m$, $\omega_b$, and $\ell_A$ to the WMAP observed values and
proceed as before.

Since the sound horizon $s$ in Eqn.~(\ref{eqn:SoundHorizon}) is
evaluated at $z>1000$, its value is unaffected by Cardassian
modifications which are important only at late times.  Hence $s$ is
essentially independent of $q$ and $n$; additionally it has only a
minor dependence on $h$.  For a given $\omega_b$ and $\omega_m$, $s$
varies by less than 0.1\% over the entire allowed parameter space.
However, the distance to the surface of last scattering $D$ is
evaluated over a period when the Cardassian modifications are
important so that $D$ has a non-trivial dependence on $q$ and $n$.

Since the CMB peaks are generated prior to and during the time of last
scattering, when the usual Friedmann approximation holds, $\ell_A$
remains the same, independent of the Cardassian parameters $q$ and $n$.
Hence $\ell_A$ is fixed by the WMAP data, having the same
value in the Cardassian model as in the standard Friedmann case.
Due to Eqn.~(\ref{eqn:AcousticScale}), the fact that $\ell_A$ and $s$ are
unchanged relative to the standard $\Lambda$ model constrains the
value of $D$ to be unchanged as well.  However, since $D$ depends
non-trivially on $q$, $n$, and $h$, as described in the previous
paragraph, we must allow all three parameters to vary.  In particular,
the value of $h$ must be allowed to vary in such a way that $D$
remains constant; i.e.,
\begin{equation} \label{eqn:ChFct}
 h \equiv h ( \omega_m, \omega_b, \ell_A, q, n )
\end{equation}
That is, for WMAP observed values for $\omega_m$, $\omega_b$, and
$\ell_A$, and for chosen values of the parameters $q$ and $n$,
numerical routines can be used to find the value of $h$ necessary to
satisfy $\ell_A = \frac{\pi D}{s}$.  The chosen parameters $q$ and
$n$ and the numerical solution for $h$ may then be used to determine
the age of the universe using Eqn.~(\ref{eqn:CardAge}).

\section{\label{sec:Results} Results}

\begin{table}    
  \centering
  \begin{tabular*}%
      {\lesserofdim{0.60\textwidth}{1.00\columnwidth}}%
      { c @{\extracolsep{\fill}} c c c }
    \hline \hline
    \noalign{\vspace{0.5 ex}}
    $q$    & $n$   & $h$             & $t_0$ (Gyr)    \\[0.5 ex]
    \hline
    \noalign{\vspace{0.5 ex}}
    $1$    & $0$   & $0.71 \pm 0.04$ & $13.7 \pm 0.2$ \\
    $1.5$  & $0.2$ & $0.72 \pm 0.04$ & $13.6 \pm 0.2$ \\
    $2$    & $0.3$ & $0.72 \pm 0.03$ & $13.6 \pm 0.2$ \\
    $10$   & $0.4$ & $0.76 \pm 0.03$ & $13.4 \pm 0.2$ \\
    $100$  & $0.4$ & $0.77 \pm 0.03$ & $13.4 \pm 0.2$ \\[0.5 ex]
    \hline \hline
  \end{tabular*}
  \caption[Cardassian Parameter Cases]{
    Values for the Hubble constant $h$ and age of the universe $t_0$
    for selected SNe~Ia allowed parameters in the Cardassian model
    \cite{Wang:2003cs}.
  }
  \label{table:Cases}
\end{table}

\begin{figure}    
  \includegraphics[width=\lesserofdim{0.80\textwidth}{0.95\columnwidth}]{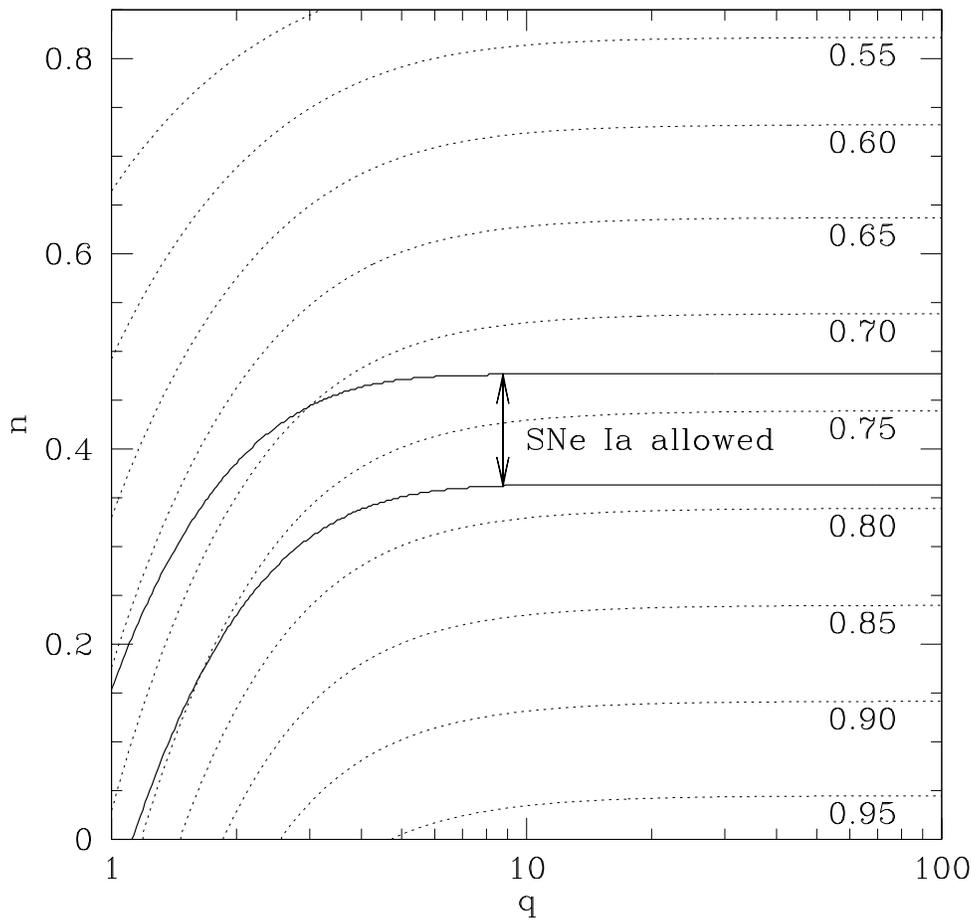}
  \caption[Hubble Constant Contour Plot]{
    Hubble constant $h$ contours (in units of $100$ km/s/Mpc) for
    various values of $q$ and $n$ in the MP Cardassian model (defined
    in Eq.(\ref{eqn:Cardassian})), based upon central values of WMAP
    data.  Uncertainties associated with several cases may be found in
    Table~\ref{table:Cases}.  The solid lines show the supernovae Ia
    allowed region of parameter space determined by Wang \textit{et al}.\
    \cite{Wang:2003cs}.  }
  \label{fig:hContour}
\end{figure}

\begin{figure}
  \includegraphics[width=\lesserofdim{0.80\textwidth}{0.95\columnwidth}]{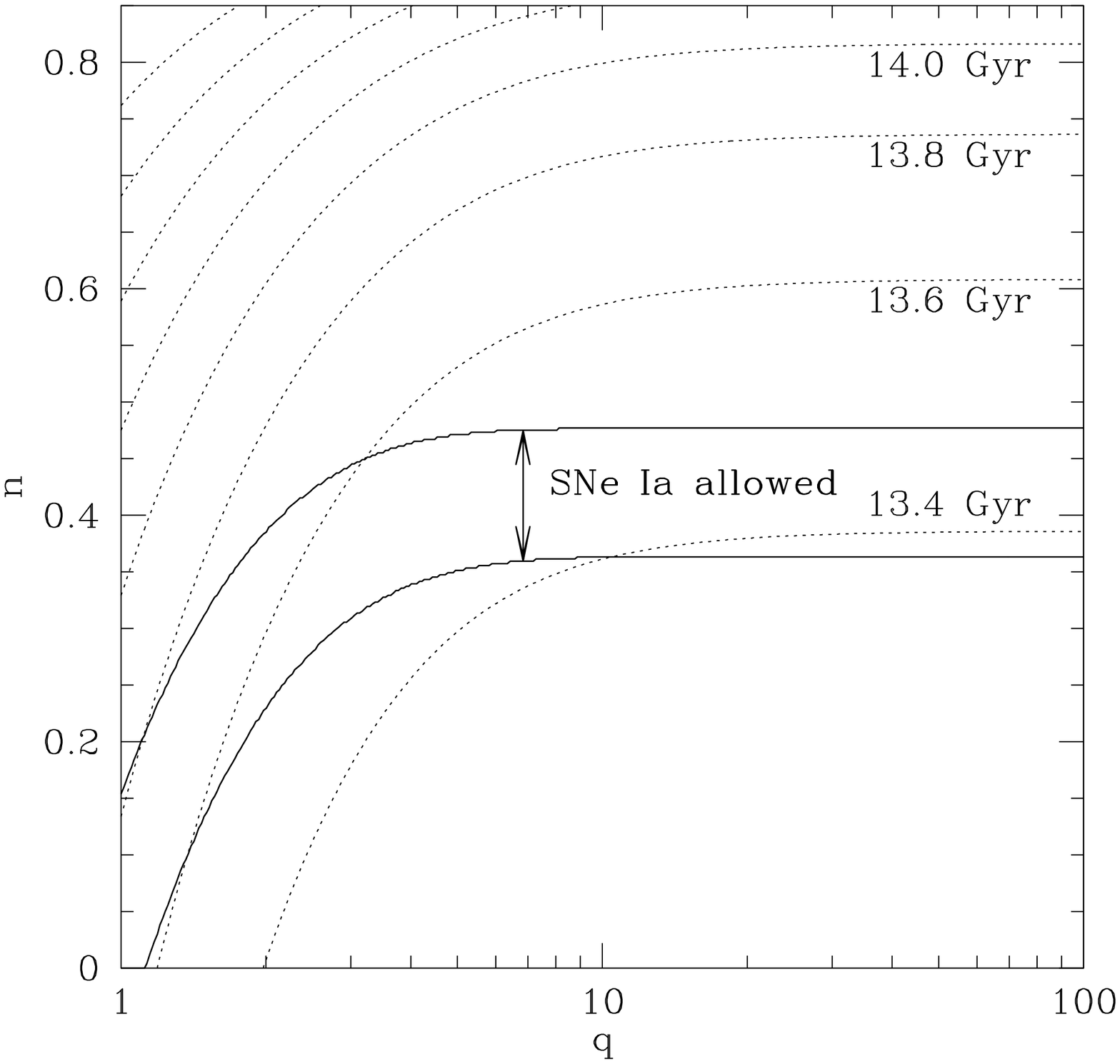}
  \caption[Age of the Universe Contour Plot]{
    Age of the universe contours (in Gyrs) for various values of $q$
    and $n$ in the MP Cardassian model (defined in
    Eq.(\ref{eqn:Cardassian})), based upon central values of WMAP data.
    Uncertainties and SNe~Ia allowed region  are as described in
    Figure~\ref{fig:hContour}.  }
  \label{fig:ageContour}
\end{figure}

We have obtained the age of the universe appropriate to the MP
Cardassian modification to the Friedmann equation given in
Eq.~(\ref{eqn:Cardassian}).  Our results are displayed in Figures
\ref{fig:hContour} and \ref{fig:ageContour}.
Since the best fit value of the Hubble constant now depends on the
new parameters $n$ and $q$, we present a contour plot of $h$ over this
parameter space in Figure~\ref{fig:hContour}.  A contour plot of the
age is shown in Figure~\ref{fig:ageContour}.  The contours were
generated using the central values for $\omega_m$, $\omega_b$, and
$\ell_A$ observed by WMAP; uncertainties for several choices of the
parameters are shown in Table~\ref{table:Cases}.

Constraints from SNe~Ia data on the free parameters $q$ and $n$, which
were obtained  by Wang \textit{et al.}\ \cite{Wang:2003cs}, are
illustrated in both the $h$ and age contour plots.
These constraints were derived using fixed $\Omega_{0m} = 0.3$;
recall, however, that we have fixed $\omega_m$ in our analysis, in which
case $\Omega_{0m}$ varies.  It would likewise be more appropriate to fix
$\omega_m$ for deriving SNe~Ia constraints.  The constraints in this
case, though, would not significantly differ.  For $n \approx 0$ and
$q \approx 1$, the MP Cardassian model is nearly equivalent to the
$\Lambda$-CDM model and $\Omega_{0m} = 0.3$ is valid; hence the
SNe~Ia constraints in this region should remain valid.  For $q \gg 1$,
the luminosity distances to the supernovae are essentially independent
of $\Omega_{0m}$ and the constraints shown in this region are also
valid
\footnote{
From Eqns.~(\ref{eqn:CHubbleE})-(\ref{eqn:CfX}), one can show for
$q \gg 1$ that $E_C(z)^2 \to (1+z)^{3n}$, independent of $\Omega_{0m}$,
when $\Omega_{0m} (1+z)^{3(1-n)} < 1$.  For the $z<1$ redshifts of the
SNe~Ia measurements and $n$ \& $\Omega_{0m}$ values of the SNe~Ia
constraints, the above constraint is satisfied (note the above limit
is \textit{not} valid for much larger $z$).  The luminosity distance
to the SNe, $d_L(z) = (1+z) \int_0^z dz^\prime\,[E(z^\prime)]^{-1}$,
is therefore independent of $\Omega_{0m}$.
}.
With the fixed $\Omega_{0m}$ constraints valid for $q \approx 1$
and $q \gg 1$, the intermediate $q$ constraints are unlikely to be
significantly different from the fixed $\omega_m$ values (the
constraints would otherwise have to exhibit fairly pathological
behavior).  So while the SNe~Ia constraints shown here are not
rigorously correct for our analysis, they are fairly close to the
``true'' constraints and should still be valid.

Other comparisons of the MP Cardassian model to observational data have
also been done \cite{Wang:2003cs,Amarzguioui:2004kc,
Alcaniz:2005ki,Koivisto:2004ne,Gong:2003st,Gong:2004sa,Nesseris:2004wj}.
In addition to the SNe~Ia data, Wang \textit{et al.}\  determined
constraints on the $q$-$n$ parameter space from the shift in the CMB
angular spectrum \cite{Wang:2003cs}; these constraints are weaker than
those of the SNe~Ia data, except at $q<2$, where the upper bound on $n$
is somewhat reduced (see Figure 2 of their paper).
Amarzguioui \textit{et al.}\ have combined CMB and SNe~Ia measurements
to determine constraints on the parameter space very similar to that
of the Wang \textit{et al.}\ SNe~Ia constraints, but with a preference
for lower values of $q$ \cite{Amarzguioui:2004kc} (see Figure 3 of
their paper; the $1\sigma$ contour gives $q \lae 4$).
Note, by including the CMB spectrum with the SNe~Ia observations,
the concerns over the validity of the Wang \textit{et al.}\ limits
discussed in the previous paragraph do not apply to the limits given by
Amarzguioui \textit{et al.}

Values for $h$ and $t_0$ are shown in Table~\ref{table:Cases} for
several cases of $q$ and $n$ lying within the SNe~Ia allowed region.
The case $q = 1$ \& $n = 0$, which is equivalent to the $\Lambda$-CDM
model, reproduces  the WMAP results with a Hubble parameter $h$ of
$0.71 \pm 0.04$ and an age of $13.7 \pm 0.2$ Gyr.

Even over the entire parameter space we considered for the
MP Cardassian model (with $q$ ranging from 1-100), the age varies
very little, ranging from $13.2$ to $14.6$ Gyr.  The SNe~Ia
allowed band roughly follows the contours and contains values
ranging from $13.4$ to $13.8$ Gyr.  This range is consistent with
other constraints on the age
\cite{Age:GlobularClusters,Age:Binaries,Age:WhiteDwarfs,Age:Radioactive,
Cayrel:2001qi};
in particular, it is above the minimum of $t_0 > 12.6_{-2.4}^{+3.4}$
Gyr determined from globular cluster ages by Krauss \& Chaboyer
\cite{Krauss:2003yb} as well as a minimum of $t_0 > 12.5 \pm 3$
Gyr determined from radioisotope studies by Cayrel \textit{et al.}
\cite{Cayrel:2001qi}.

While the CMB determination of $h$ depends on the overall model, other
techniques measure its value more directly.  The Hubble Space
Telescope (HST) Key Project analyzed Cepheids over distances of 60 to
400 Mpc to determine a value of $0.72 \pm 0.08$ \cite{Freedman:2000cf}.
Measurements over such short distances are not significantly affected
by the variations in the cosmological model.  While the value of $h$
obtained in Figure 1 does vary significantly over the entire allowed
($n < 2/3$) parameter space, ranging from $0.50$ to $0.97$, we can see
that the SNe~Ia allowed band roughly follows the contours and contains
values between $0.66$ and $0.78$.  This allowed band is thus
consistent with the HST results.  Thus we find the MP Cardassian model
to be in general agreement with other independent determinations and
limits for the Hubble constant and age of the universe.

We now have estimates of the matter density in these models as well.
From the value of $\omega_m$ measured by the CMB (given in
Table~\ref{table:Cases}), together with the estimates of the Hubble
constant given in Figure~\ref{fig:hContour}, we can extract $\Omega_m$.
For the full $n < 2/3$ parameter space discussed in the last paragraph,
we find $0.14 \leq \Omega_{0m} \leq 0.54$.  The range is far more
restricted if we consider only those values consistent with the SNe~Ia
allowed band: $0.22 \leq \Omega_{0m} \leq 0.31$;
this result is in agreement with that of Amarzguioui \textit{et al.}\ 
\cite{Amarzguioui:2004kc} ($\Omega_{0m} \sim 0.3$ in their analysis).

In summary, we have shown how to determine the age of the universe in a
MP Cardassian model from CMB data, which requires the appropriate choice
of CMB derived parameters.  We find that the CMB implied age is entirely
consistent with other observational constraints for essentially the
entire MP Cardassian parameter space examined here (alternatively,
observational constraints on the age do not constrain the Cardassian
parameter space).  When restricting the parameter space to that
consistent with SNe~Ia measurements \cite{Wang:2003cs}, the age of the
universe falls somewhere between 13.4 and 13.8 Gyr, a range that
includes the age in a $\Lambda$-CDM model.


\begin{acknowledgments}
  CS thanks D. Spergel for helpful conversations, particularly for
  suggesting the appropriate orthogonal parameters.  We also thank
  Y. Wang for useful discussions.  We acknowledge the support of the
  DOE and the Michigan Center for Theoretical Physics via the
  University of Michigan.
\end{acknowledgments}




\end{document}
